# The Doppler Peaks from Cosmic Texture

R.G. Crittenden and Neil Turok

*Joseph Henry Laboratory, Princeton University,*

*Princeton NJ, 08544.*

(5/25/95)

## Abstract

We compute the angular power spectrum of temperature anisotropies on the microwave sky in the cosmic texture theory, with standard recombination assumed. The spectrum shows 'Doppler' peaks analogous to those in scenarios based on primordial adiabatic fluctuations such as 'standard CDM', but at quite different angular scales. There appear to be excellent prospects for using this as a discriminant between inflationary and cosmic defect theories.

The cosmic microwave background (CMB) anisotropy is the cleanest probe we have of structure formation in the universe, providing our best hope for definitively confirming or excluding theories of the origin of structure. The COBE observation [1] provided the first solid evidence of anisotropy, but due to its limited angular resolution it provided little more than an overall normalisation and rough evidence of scale invariance. Higher resolution experiments offer the far richer prospect of a detailed measurement of the angular power spectrum and statistical properties of the anisotropies, which will give very powerful constraints on theories.

Existing theories of the origin of structure fall into two categories, based upon either the amplification of quantum fluctuations in a scalar field during inflation or upon the process of symmetry breaking and field ordering. The latter category includes the theories of cosmic strings, global monopoles and textures. While techniques for computing density perturbations in inflationary theories are well established [2,3], greater uncertainty is attached to the field ordering theories because they are inherently nonlinear and nonGaussian.

Nevertheless there has been progress, particularly in those theories for which the nonlinear sigma model describes the field ordering process [4]. Recently the degree scale CMB anisotropies were calculated for the cosmic defect theories under the simplifying assumption that the universe was fully ionised [5]. Reionisation is more likely in defect theories than in Gaussian theories because large density perturbations around the defects could cause early star formation, releasing ionising radiation. But the extent of this effect is uncertain because it depends on the efficiency of star formation, which is poorly understood.

In this Letter we compute the power spectrum of CMB anisotropies in the texture theory assuming instead standard recombination. In particular, we wish to see whether acoustic oscillations of the photon-baryon-electron (PBE) fluid produce 'Doppler peaks' analogous to those in inflationary theories. The calculation is harder than in the reionised case, but simplifies if we concentrate on the power spectrum alone rather than attempting to make sky maps. While the nonGaussianity of such maps is undoubtedly a key signature for cosmic defects, the CMB anisotropy power spectrum is likely to be measured first and that is what we focus on here.

The power spectra of perturbations depend only on the two-point correlations of the field stress energy tensor, $\Theta_{\mu\nu}$, and these one can hope to model in a simple way. We use the three dimensional field simulations to construct a model for the $\Theta_{\mu\nu}$ correlations, which are then fed into linear perturbation theory codes to compute density perturbations and CMB anisotropy spectra. As a check of our model, we perform full 3d fluid simulations of the source fields, gravity and matter in the tight coupling epoch and compare the perturbation spectra and cross correlations in the simulations with those predicted by the model.

There are a number of advantages to this hybrid approach. A single 3d simulation has very limited dynamic range, but many such simulations can be combined to help construct a model valid over all relevant length scales. Such a model allows one to calculate ensemble averaged power spectra without the noise inherent in 3d simulations. It can also be used in a Boltzmann calculation to compute the anisotropies for arbitrary ionisation histories. Finally, the model may provide some insight into the physics behind the generation of perturbations.

In field ordering theories, the simplest assumption is that the universe began in a homogeneous and isotropic initial state. A symmetry breaking phase transition occurs, during which some field with 'angular' degrees of freedom is given a nonzero vacuum expectation value, $\vec{\phi}^2 = \phi_0^2$. The source for perturbations is $\Theta_{\mu\nu}$, the stress energy tensor of the ordering fields, interacting with the matter and radiation only gravitationally. Since the field ordering process is causal, it follows that the correlations of the fluctuating part of $\Theta_{\mu\nu}$ are strictly zero for spacetime points whose past light cones extrapolated back to the time of the phase transition do not overlap.

We work in synchronous gauge, setting all perturbation variables zero before the symmetry breaking phase transition. The metric and matter perturbation variables $h_{ij}$ and $\delta_N$ obey causal evolution equations. Since the power spectrum is the Fourier transform of the correlation function, it follows that all these variables have 'white noise' power spectra on superhorizon scales at all times.

The source fields interact gravitationally, as described by the linearised Einstein equations. These are simplified by decomposing them into scalar, vector and tensor pieces: in Fourier space, a symmetric tensor, $T_{ij}(x) = \Sigma_\mathbf{k} T_{ij}(\mathbf{k}) e^{i\mathbf{k}\cdot\mathbf{x}}$, can be written as

$$T_{ij}(\mathbf{k}) = \frac{1}{3}\delta_{ij}T + (\hat{k}_i\hat{k}_j - \frac{1}{3}\delta_{ij})T^S + (\hat{k}_i T_j^V + \hat{k}_j T_i^V) + T_{ij}^T \qquad (1)$$

where $T_i^V k_i = k_i T_{ij}^T = T_{ij}^T k_j = T_{ii}^T = 0$. A complete set of evolution equations for the metric perturbation variables is:

$$\frac{\dot{a}}{a}\dot{h} + \frac{k^2}{3}h^- = 8\pi G a^2 \sum_N \rho_N \delta_N + 8\pi G \Theta_{00} \qquad (2)$$

$$k^2 \dot{h}^- = -24\pi G i k \sum_N (p_N + \rho_N) v_N + 24\pi G \Pi \qquad (3)$$

$$\ddot{h}^T_{ij} + 2\frac{\dot{a}}{a}\dot{h}^T_{ij} + k^2 h^T_{ij} = 16\pi G \Theta^T_{ij} \tag{4}$$

$$\ddot{h}^V_i + 2\frac{\dot{a}}{a}\dot{h}^V_i = 16\pi G \Theta^V_i. \tag{5}$$

where $\Pi \equiv \partial_i \Theta_{0i}$. The metric perturbations are defined by $ds^2 = a^2(\tau)(-d\tau^2 + (\delta_{ij} + h_{ij}(x,\tau))dx^i dx^j)$, with $\tau$ conformal time and $a(\tau)$ the scale factor, and $k^2 h^- \equiv k^2(h - h^S)$ is proportional to the scalar curvature of the constant $\tau$ spatial slices. The variables $\rho_N$, $p_N$, $\delta_N$ and $v_N$ refer as usual to the density, pressure, fractional density perturbation and velocity of each fluid - photons, baryons, cold dark matter and neutrinos. In the tightly coupled regime, the equations for the fractional perturbation in the photon density $\delta_\gamma$ are

$$\dot{\delta}_\gamma = -\frac{2}{3}\dot{h} + \frac{4}{3}ikv_\gamma; \qquad \dot{v}_\gamma = \frac{3}{4}ikc_s^2\delta_\gamma - \frac{\dot{a}}{a}(1-3c_s^2)v_\gamma \tag{6}$$

where $c_s^2 = \frac{1}{3}\rho_\gamma/(\rho_\gamma + \rho_B)$ is the square of the speed of sound, $\rho_B$ the baryon density. We model the neutrinos as a fluid with viscosity.

In the 'stiff' approximation [6], $\Theta_{\mu\nu}$ is covariantly conserved with respect to the background metric:

$$\dot{\Theta}_{00} + \frac{\dot{a}}{a}(\Theta_{00} + \Theta) = \Pi; \qquad \dot{\Pi} + 2\frac{\dot{a}}{a}\Pi = -\frac{k^2}{3}(\Theta + 2\Theta^S). \tag{7}$$

We would like to model the behaviour of $\Theta_{\mu\nu}$ while respecting these constraints. The correct underlying dynamics is that of scalar fields obeying the nonlinear sigma model equation, but a crude description is that spatial gradients in the scalar fields are 'frozen' outside the horizon, and redshift away after horizon crossing. This may be viewed as a change in the effective equation of state: if there were only spatial gradients outside the horizon, we would have $\Theta = -\Theta_{00}$. If these gradients redshifted away like radiation we would have $\Theta_{00} = +\Theta$ inside the horizon. This rough picture forms the basis of our model.

In our model we treat the source as a fluid with a scale-dependent equation of state: $\Theta = \gamma(k,\tau)\Theta_{00}$ and $\Theta^S = \gamma^S(k,\tau)\Theta_{00}$, where $\gamma$ and $\gamma^S$ are determined from field simulations. In the pure matter or radiation epochs they have scaling form, being functions only of $k\tau$. With $\gamma$ and $\gamma^S$ fixed, equations (7) determine the evolution of all scalar parts of $\Theta_{\mu\nu}$.

The main simplification this model makes is that the evolution equations for $\Theta_{\mu\nu}$ are linear. While a given mode is outside the horizon it quickly settles into a 'scaling' solution specified by a single amplitude. It follows that in the long time limit unequal time cross correlations factorise, i.e.

$$\langle A_{\mathbf{k}}(\tau) B_{-\mathbf{k}}(\tau') \rangle = \langle A_{-\mathbf{k}}(\tau) B_{\mathbf{k}}(\tau') \rangle = a(k,\tau) b(k,\tau') \tag{8}$$

where $a(k,\tau)$ and $b(k,\tau)$ are 'master' functions satisfying the same linear equation that the random field modes $A_{\mathbf{k}}$ and $B_{\mathbf{k}}$ do. Setting $B_{\mathbf{k}} = A_{\mathbf{k}}$ and $\tau = \tau'$, one sees that $a^2(k,\tau)$ is just the power spectrum of $A_{\mathbf{k}}(\tau)$. The functions $\gamma$ and $\gamma^S$ are given by $\langle \Theta_{00} \Theta \rangle / \langle \Theta_{00}^2 \rangle$ and $\langle \Theta_{00} \Theta^S \rangle / \langle \Theta_{00}^2 \rangle$, which we measure in field simulations and perform a simple fit [7] to specify the model.

The initial conditions for the model are determined by scaling and causality. Scaling implies that all correlators should be specified in terms of a single scale $\tau$ alone. Dimensional analysis then gives

$$\langle |\Theta_{00}(\mathbf{k}, \tau)|^2 \rangle = \frac{f(k\tau) \phi_0^4}{V\tau} \tag{9}$$

where $V$ is a fiducial comoving volume. Causality implies a white noise spectrum for $\Theta_{00}$, so that $f(k\tau) = $ constant for $k\tau \ll 1$. Thus for each $k$ the master function for $\Theta_{00}$ should be by a fixed constant times $\tau^{-\frac{1}{2}}$ at early times. Scaling of $\Theta$ and equation (7) fix $\gamma(0) = -\frac{1}{2}$ in the radiation era.

Let us make two comments on the most obvious limitations of the model. Examination of the large N expressions [8] shows that the Fourier modes of the energy momentum tensor far outside the horizon ($k\tau \ll 1$), or far inside the horizon ($k\tau \gg 1$) are dominated by interference terms involving horizon wavelength modes, where the most important dynamics is taking place. These interference effects are unlikely to be well represented by the dynamics of a fluid in which all Fourier modes decouple. However, the dominant perturbations in a given Fourier mode of the metric and matter variables are produced by the source around horizon crossing, so all we really require is that the model adequately represents the source at this time. At horizon crossing, only dynamics on a single length scale are involved, and these may be reasonably well described by fluid equations.

A second concern regards unequal time correlations. Since the perturbation equations are linear, the solution for the set of perturbation vartiables $\delta_a$ takes the form $\delta_a(k,\tau) = \int_0^\tau d\tau' G_{ab}(k,\tau,\tau') S_b(k,\tau')$, where $S_b$ are the sources $\Theta_{00}$, $\Pi$, and $G_{ab}$ the appropriate Greens functions. It follows that all power spectra of the perturbations are determined by the *unequal* time source correlators. Equation (8) shows that the model actually builds in *maximal* unequal time correlations, leading to an overestimate of the coherence in time of the source terms [9]. We argue again, however, that around horizon crossing there is only one length and timescale involved, and the timescale for decoherence of the source is of the same order as that for the source to redshift away. This is what our model builds in. Nevertheless it is of some importance to check the model predictions against a full 3d simulation to see whether the predicted coherent oscillations are really there.

Figure 1 shows the results of a 3d simulation of equations (2 -5) sourced by texture fields. The Figure shows the power in the variables $h^-$, $\delta_\gamma$, $\delta_C$, $v_\gamma$ from the 3d code against those predicted by model, at the instant of matter-radiation decoupling. There are no adjustable parameters in the comparison, and there is good qualitative agreement between the simulations and the model. We have similarly checked the power spectra of the source terms $\Theta_{00}$ and $\Pi$ and found reasonable agreement. As mentioned, we are particularly concerned to check whether the radiation oscillations evident in the model are really present, since these give rise to the Doppler peaks. The power spectrum for $\delta_\gamma$ shows some evidence of coherent oscillations, but of small amplitude. The power spectrum of $v_\gamma$ shows larger oscillations. The difference is explained by looking at the behaviour of the model. Whereas the master function for $\delta_\gamma$ oscillates about zero, with neighbouring peaks of similar amplitude, $v_\gamma$ oscillates about a constant offset, so that neighbouring peaks have quite different amplitude. Smearing effects, whether from numerical effects or from real incoherence in the perturbations, do less to reduce the amplitude of these oscillations.

A difficulty in computing power spectra from the 3d simulations is that oscillations in power are smeared out by sampling effects and by noise, which prevent one ever seeing a zero in a positive definite quantity like the power. A cleaner check of whether phase-coherent oscillations are present in the 3d simulations is to measure the cross correlation of $\delta_\gamma$ with some variable that does *not* have oscillatory behavior, such as $h^-$. Figure 2 shows $\langle \delta_\gamma h^- \rangle$ measured in the simulations compared with the model predictions. If the oscillations were incoherent this cross correlation would average to zero. Similarly, we have checked $\langle \delta_C h^- \rangle$ and confirmed that a single sign change takes place just as the model predicts.

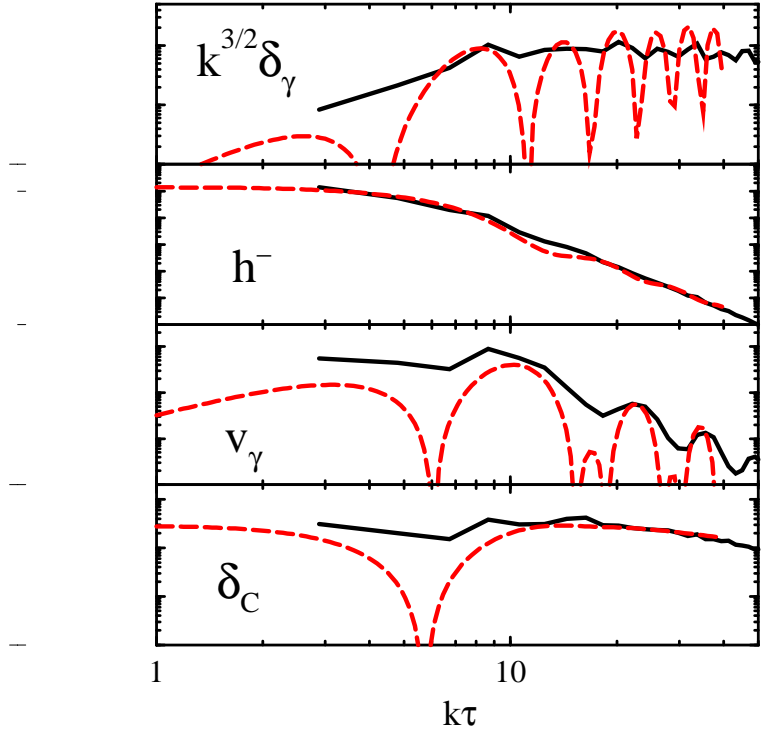

FIG. 1. Power spectra from three dimensional simulations (solid lines) compared to those predicted by the equation of state model (dashed lines). The perturbation variables are those defined in equations (2-6). The vertical scales are arbitrary: large ticks are separated by an order of magnitude.

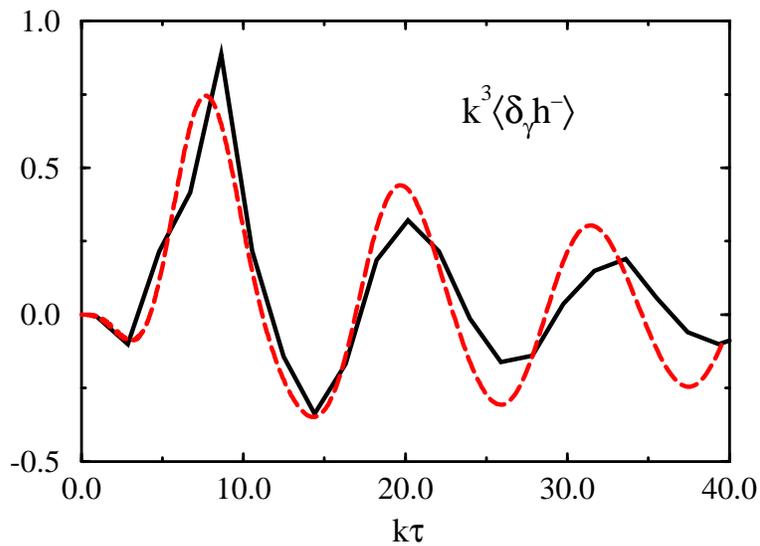

FIG. 2. The cross correlation $k^3 \langle \delta_\gamma(k) h^-(k) \rangle$ measured in a single simulation (solid) is compared to the predictions of our model (dashed line). This provides strong evidence that coherent oscillations exist in the photon baryon fluid.

All this we take as evidence that the scalar part of the model is working well. The large scale microwave anisotropies also involve vector and tensor perturbations and we have treated these in the same 'factorisation' approximation. That is, we compute the power spectrum for $\Theta^V$ and $\Theta^T$ in 3d simulations and approximate the unequal time correlation functions as products of the square roots of the relevant power spectra. As mentioned above, it is hard to see oscillations in the true power spectra from those measured in the 3d simulations. To check our sensitivity to this effect we have also modelled the vector and tensor cross correlations using oscillatory 'master functions' whose envelope has the same amplitude. The change alters the vector and tensor power spectra somewhat, but has a small effect on our final multipole power spectrum.

In the approximation of instantaneous recombination one can write the full expression for the microwave anisotropy in a direction $\mathbf{n}$ as the sum of 'intrinsic', 'Doppler' and 'integrated Sachs-Wolfe' pieces,

$$\frac{\delta T}{T}(\mathbf{n}) \equiv \sum_{l,m} a_{lm} Y_{lm}(\theta, \phi) = \frac{1}{4}\delta_\gamma - \mathbf{v}_\gamma \cdot \mathbf{n} - \frac{1}{2}\int_i^f d\tau \dot{h}_{ij} n^i n^j \qquad (10)$$

Converting this into a spectrum of $C_l's$ involves ensemble averaging which is simple in the factorisation approximation. The scalar expression is well known (see e.g. [10]), the vector and tensor contributions are even simpler. We define the vector master function by $\langle \dot{h}_i^V(\mathbf{k}) \dot{h}_j^V(-\mathbf{k}) \rangle = \dot{h}^V(k,\tau) \dot{h}^V(k,\tau')(\delta_{ij} - \hat{k}_i \hat{k}_j)$, and the tensor function by $\langle \dot{h}_{ij}^T(\mathbf{k}) \dot{h}_{kl}^T(-\mathbf{k}) \rangle = \dot{h}^T(k,\tau) \dot{h}^T(k,\tau')(\delta_{ik}\delta_{jl} + ...)$ where ... denotes terms imposed by the transverse-traceless and symmetry conditions. Some algebra then shows that

$$\langle|a_{lm}^V|^2\rangle = 16\pi^2 \int \frac{k^2 dk}{(2\pi)^3} l(l+1)(\int \frac{d\tau}{k\tau} j_l'(k\tau)\dot{h}^V(k,\tau))^2 \qquad (11)$$

$$\langle|a_{lm}^T|^2\rangle = 4\pi^2 \int \frac{k^2 dk}{(2\pi)^3} \frac{(l+2)!}{(l-2)!}(\int \frac{d\tau}{(k\tau)^2} j_l(k\tau)\dot{h}^T(k,\tau))^2. \qquad (12)$$

The latter equation agrees with (and simplifies) the formula given by Abbot and Wise [11]. The functions $h^V$ and $h^T$ are obtained by solving the Einstein equations (4) and (5) numerically, using the master functions for $\Theta^V$ and $\Theta^T$ inferred from 3d simulations.

The final CMB anisotropy power spectrum is presented in Figure 3, along with various components. The results shown were computed using a fluid code, in the 'instantaneous recombination' approximation. We have checked the scalar and tensor calculations using a full Boltzmann code, with very good agreement. The cosmological parameters used were $h = 0.5$, $\Omega = 1$, $\Omega_B = 0.05$.

A word about uncertainties. We have found the *shape* of the spectrum for $l > 100$ to be remarkably insensitive to the details of our model: plausible variations change the results by less than 10%. Likewise the qualitative shape of the vector and tensor contibutions is fairly model-independent. We attach greater uncertainty (of order 50%) to amplitude of the perturbations on large (COBE) scales, first because of the complexity of extracting the vector and tensor master functions from the 3d code, discussed above, and second because the scalar power on these scales *is* sensitive to variations in the 'equations of state' as one passes from the radiation to the matter era. More detailed numerical studies will resolve these uncertainties.

The most striking difference between the texture and inflationary spectra is that 'Doppler' peaks and troughs are almost 'out of phase' - where inflation predicts a maximum texture gives a minimum, and vice versa. This behaviour is reminiscent of that found in 'isocurvature' models [12], and it occurs for a similar reason [13]. The radiation oscillations are driven by the metric perturbations: from equations (6) we have $\ddot{\delta}_\gamma + c_S^2 \delta_\gamma = -\frac{2}{3}\ddot{h}$. The phase of the oscillations is determined by the behaviour of $\ddot{h}$ as a mode crosses the horizon. In 'adiabatic' theories like the simplest inflationary models, one has superhorizon curvature perturbations, in which the metric perturbation $h \propto \tau^2$. Thus the force driving for the radiation oscillations is to a first approximation constant. In the 'isocurvature' theories, the source term is not constant and this leads to a relative phase shift in the oscillations.

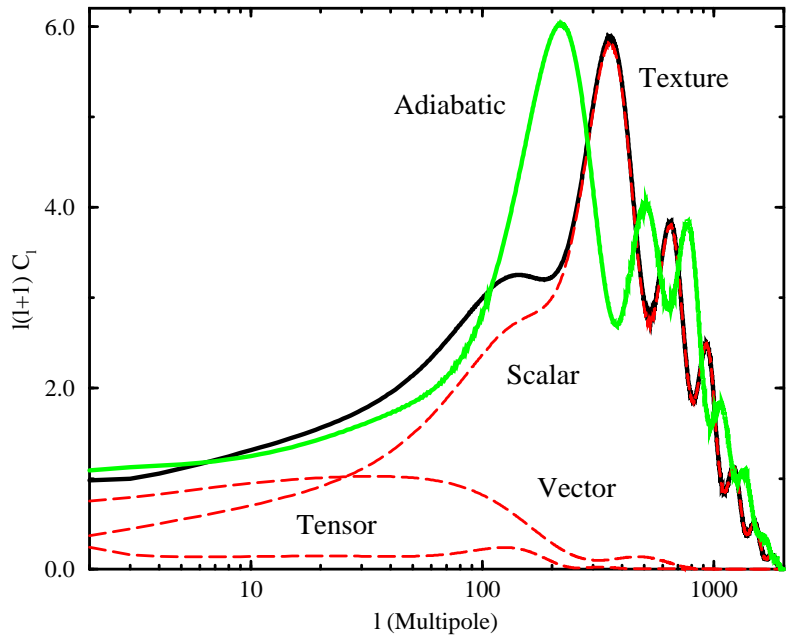

FIG. 3. The predicted anisotropy power spectrum $C_l = \langle |a_{lm}|^2 \rangle$ (equation (11)), decomposed into scalar, vector and tensor contributions. The vector and tensor parts have been multiplied by a factor of two for clarity. Shown for comparison (grey curve) is the prediction of the simplest inflationary theory with cold dark matter with the same cosmological parameters, $h = 0.5$, $\Omega_B = 0.05$, and $\Omega = 1$.

The field ordering theories are *both* adiabatic and isocurvature: there are no physical perturbations in the space curvature or entropy on superhorizon scales. This leads to a different behavior of the $\ddot{h}$ forcing term around horizon crossing. Scale invariance and dimensional analysis imply that

$$\langle |h(\mathbf{k})|^2 \rangle = \frac{g(k\tau)\tau^3}{V} \qquad (13)$$

with $g$ a dimensionless function. In standard 'adiabatic' theories, $h \propto \tau^2$ so $g \propto k\tau$ for small $k\tau$. But for field ordering theories, the small $k$ behaviour of $g$ is fixed instead by causality: $h$ must have a 'white noise' power spectrum, and so $g =$ constant and thus $h \propto \tau^{\frac{3}{2}}$ at small $k\tau$. Thus the forcing term decreases as $\tau^{-\frac{1}{2}}$. It follows that $\delta_\gamma$ reaches its first maximum sooner than in the adiabatic theories. It is intriguing that, at least in the context of adiabatic theories, the large scale angular power spectrum, and in particular the phase of the 'Doppler' peaks could be telling us something as fundamental as whether the perturbations were generated causally within the standard big bang, or were necessarily generated in a preceding inflationary epoch.

We thank D. Coulson for collaboration in development of the 3d code. This work was partially supported by NSF contract PHY90-21984, and the David and Lucile Packard Foundation.